\documentclass[aps,pra,twocolumn,superscriptaddress,showpacs]{revtex4}
\usepackage{amsmath,amssymb,graphicx}
\usepackage[hypertex,colorlinks=true,citecolor=blue,urlcolor=blue]{hyperref}

\newcommand{\PR}[4]{
Phys. Rev. #1 {\bf #2}, \href{http://dx.doi.org/10.1103/PhysRev#1.#2.#3}{#3} (#4)}
\newcommand{\PRRC}[4]{
Phys. Rev. #1 {\bf #2}, \href{http://dx.doi.org/10.1103/PhysRev#1.#2.#3}{#3(R)} (#4)}
\newcommand{\PRL}[3]{
Phys. Rev. Lett. {\bf #1}, \href{http://dx.doi.org/10.1103/PhysRevLett.#1.#2}{#2} (#3)}
\newcommand{\RMP}[3]{
Rev. Mod. Phys. {\bf #1}, \href{http://dx.doi.org/10.1103/RevModPhys.#1.#2}{#2} (#3)}
\newcommand{\JPSJ}[3]{
J. Phys. Soc. Jpn. {\bf #1}, \href{http://dx.doi.org/10.1143/JPSJ.#1.#2}{#2} (#3)}
\newcommand{\Nature}[4]{
Nature (London) {\bf #2}, \href{http://dx.doi.org/10.1038/#1}{#3} (#4)}
\newcommand{\Nat}[5]{
Nat. #2 {\bf #3}, \href{http://dx.doi.org/10.1038/#1}{#4} (#5)}

\begin{document}
\title{Laser-induced magnetization curve}
\author{Shintaro Takayoshi}
\affiliation{National Institute for Materials Science, 
Tsukuba, Ibaraki 305-0044, Japan}
\author{Masahiro Sato}
\affiliation{Department of Physics and Mathematics, Aoyama-Gakuin University, 
Sagamihara, Kanagawa 252-5258, Japan}
\author{Takashi Oka}
\affiliation{Department of Applied Physics, The University of Tokyo, 
Hongo, Bunkyo, Tokyo 113-8656, Japan}

\date{\today}

\begin{abstract}
We propose an all optical ultrafast method to highly magnetize 
general quantum magnets using a circularly polarized terahertz 
laser. The key idea is to utilize a circularly polarized laser and 
its chirping. Through this method, one can obtain magnetization curves 
of a broad class of quantum magnets as a function of time 
{\it even without any static magnetic field}. 
We numerically demonstrate the laser-induced magnetization process 
in realistic quantum spin models and find a condition for the realization. 
The onset of magnetization can be described 
by a many-body version of Landau-Zener mechanism. 
In a particular model, we show that a plateau 
state with topological properties can be realized dynamically. 
\end{abstract}


\maketitle

\section{Introduction}
\label{sec:Intro}

Ultrafast control of magnetization has become a hot topic 
recently~\cite{Kimel05,Kampfrath11,Kirilyuk10,Vicario13,Miyashita95,
Miyashita98,Thomas96}. 
Not only does this technique have much potential for application, e.g.,
fast data storage and spintronics~\cite{Zutic04}, 
but it also poses an important question in fundamental physics: 
Can we coherently induce an ultrafast phase transition 
in many-body quantum systems? 
Terahertz (THz) laser~\cite{Fulop11,Hoffmann09,Hirori11,
Matsunaga13,SatoGonokami13} 
is preferred in terms of quantum coherence since its photon energy 
is comparable with the energy scale of spin systems. 
Among spin systems, quantum antiferromagnets are known to 
show rich many-body effects. A traditional way to study their properties 
is to measure the magnetization curve, i.e., relation between magnetization 
and externally applied magnetic field. 
Prominent phenomena such as various 
magnetization plateaux~\cite{Oshikawa97a,Takigawa11}, 
field induced topological states~\cite{Tanaka09,Oshikawa00},
and Bose-Einstein condensation of magnons~\cite{Nikuni00,Giamarchi08}
have been discovered.
However, the magnetization curve up to the saturated magnetization 
often requires an extremely high magnetic field. 
If we can realize the full magnetization process 
in table-top laser experiments, 
studies on nontrivial magnetic phenomena would be more accessible, 
but it is not easy due to the limited laser strength. 
The magnetic field component of a THz laser is typically less than 0.5 T,
several orders below the necessary field strength 
($\sim 10-100$ T and more) for full magnetization.

We can overcome this difficulty with help from 
recent progress of quantum many-body systems 
in time-periodic external fields, which have been studied 
both theoretically~\cite{Oka09,Kitagawa11,Lindner11,Takayoshi13,Sato14} 
and experimentally~\cite{Wang13}. 
In general quantum magnets, a proper unitary transformation 
maps a rotating magnetic field of a circularly polarized laser
into an effective static magnetic field~\cite{Takayoshi13,Ruckriegel12}.
If the magnetic field rotates in the $xy$ plane, 
the effective static field has a component in the $z$ direction,
and its strength is given by the photon energy $\Omega$.
For example, a laser with frequency of 
1 THz ($\Omega=$1 THz$\sim4$ meV) 
can typically produce an effective field as strong as $40$ T. 
Even stronger fields are obtained by increasing frequency $\Omega$, 
and it is reasonable to consider a possibility of 
``laser-induced magnetization curves''.

Our proposal is to change $\Omega$ from a small value 
to a larger value as slowly as possible. 
In other words, an upchirped THz laser pulse is required 
\cite{SatoGonokami13,Kamada13}. 
In this paper, starting from the zero field ground state (GS), 
we show that the system under a upchirped laser almost follows 
the GS in static magnetic field up to full magnetization 
in time evolution of the wave function. 
This is a significant difference compared 
with an equilibrium magnetization curve 
obtained using high magnetic field facilities 
since we can wait for a sufficiently long time 
in order to make the system equilibrated in the latter case. 
Thus, we need to find an appropriate protocol to 
realize the laser-induced magnetization curve. 
We determine ideal conditions to approach highly magnetized states 
and explain the mechanism by many-body 
Landau-Zener (LZ) tunnelings~\cite{Landau32,Zener34}.
In addition, in a particular model, 
we show that a symmetry protected topological (SPT) plateau
state~\cite{Pollmann10,Gu09} can be realized in this manner.

This paper is organized as follows. 
In Sec.~\ref{sec:BasicIdea}, we explain our idea about realizing 
laser-induced magnetization processes. 
Following the idea of Sec.~\ref{sec:BasicIdea}, 
we numerically demonstrate that a laser-induced magnetization 
indeed occurs in two simple but realistic 
quantum spin systems in Sec.~\ref{sec:LIMP}. 
It is also shown that a magnetization plateau state having 
topological properties is reproduced by circularly polarized laser. 
Sections~\ref{sec:LZtunneling} and 
\ref{sec:Topological} are devoted to the quantitative discussion about 
the laser-induced magnetization curves obtained in Sec.~\ref{sec:LIMP}. 
We consider time evolution of wave functions 
from the viewpoint of the LZ tunneling mechanism 
in Sec.~\ref{sec:LZtunneling}. 
We also obtain the ``phase diagram'' for the laser-induced plateau state 
of Sec.~\ref{sec:LIMP}, 
which leads to an ideal condition for the realization of laser-induced 
magnetization curves. 
In Sec.~\ref{sec:Topological}, we investigate how close the 
dynamically laser-induced plateau state in Sec.~\ref{sec:LIMP} is 
to the equilibrium plateau state. 
Finally we summarize and discuss our results 
in Sec.~\ref{sec:Discuss}.

\begin{figure}[t]
\includegraphics[width=0.45\textwidth]{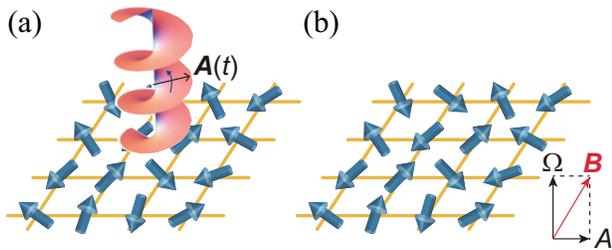}
\caption{(Color online) (a) Quantum magnet in a circularly polarized laser. 
(b) Effective static system after the unitary transform  (\ref{eq:Rotation}).} 
\label{fig:LaserMagnet}
\end{figure}

\section{Basic Idea}
\label{sec:BasicIdea}
In this section, we explain our basic idea for realization of 
laser-induced magnetization process.
We consider general quantum magnets under a 
circularly polarized laser as shown in Fig.~\ref{fig:LaserMagnet}(a). 
The magnetic field of the laser generates a  
dynamical Zeeman interaction, which results in a Hamiltonian, 
\begin{equation}
 {\cal H}(t)={\cal H}_{0}
             -A_{x}(t)S^{x}_{\rm tot}-A_{y}(t)S^{y}_{\rm tot}.
 \label{eq:HamiltonianWithLaser}
\end{equation}
The first term ${\cal H}_{0}$ is the spin Hamiltonian and 
the magnetic field of the laser 
$(A_{x},A_{y})=A(\cos(\Omega t),\sin(\Omega t))$ 
rotates in the $xy$ plane. 
The operator $S_{\rm tot}^{\alpha}$ is the $\alpha$ component 
of the total spin and $\Omega$ is the laser frequency. 
Here we assume that the spin Hamiltonian ${\cal H}_{0}$ is invariant 
under U(1) spin rotation around the $S^{z}$ axis. 
If we apply the time-dependent unitary transform 
\begin{equation}
U=\exp({\rm i}\Omega S_{\rm tot}^{z}t)
\label{eq:Rotation}
\end{equation}
to such U(1)-symmetric magnets under the laser, 
we obtain the following relation,
\begin{equation}
 U({\cal H}(t)-i\partial_{t})U^{-1}
   ={\cal H}_{0}-\Omega S_{\rm tot}^{z}
     -AS_{\rm tot}^{x}-i\partial_{t}.\nonumber
\end{equation}
This transform maps the system from the experimental frame 
to a rotating frame in spin space~\cite{Takayoshi13,Ruckriegel12}. 
In this rotating frame, the Hamiltonian becomes {\it static} as follows 
(see Fig.~\ref{fig:LaserMagnet}(b)):
\begin{equation}
 {\cal H}_{\rm eff}={\cal H}_{0}-B_{z}S^{z}_{\rm tot}
   -B_{x}S^{x}_{\rm tot}\quad 
   (B_{z}=\Omega,\; B_{x}=A).
\label{eq:EffectiveHamiltonian}
\end{equation}
We emphasize that this mapping from the dynamical system 
(\ref{eq:HamiltonianWithLaser}) to the static one 
(\ref{eq:EffectiveHamiltonian}) always holds 
when the laser is circularly polarized 
and ${\cal H}_{0}$ is U(1) symmetric around the $S^{z}$ axis. 
The mapping does not depend at all on spatial dimensions, 
spin magnitude $S$, and the other details.

Equation (\ref{eq:EffectiveHamiltonian}) indicates that 
a laser with high frequency $\Omega$ can lead to highly magnetized states 
due to the effective Zeeman term $-\Omega S_{\rm tot}^{z}$. 
It also implies that the magnetization points in the opposite 
direction by changing the helicity of the laser. 
As we mentioned in the Sec.~\ref{sec:Intro}, 
it is difficult to increase the magnitude $A$ 
to a large value, while the laser frequency $\Omega$ can be 
relatively easily tuned \cite{SatoGonokami13}. Therefore, the magnetization along the $S^{z}$ 
axis, $\langle S_{\rm tot}^{z}\rangle$, is expected to grow by 
increasing $\Omega$.

However, we should note that even if we increase $\Omega$, 
the wave function of the magnet does not always follow the GS 
of the effective static Hamiltonian (\ref{eq:EffectiveHamiltonian}). 
Instead, the system may be in the excited states and the 
magnetization may be small or even absent. 
In fact, when ${\cal H}_{0}$ has SU(2) spin-rotational symmetry, 
one can understand that the growth of the magnetization does not occur 
as follows. The effective Hamiltonian ${\cal H}_{\rm eff}$ represents 
a quantum magnet under a static magnetic field $\boldsymbol{B}=(A,0,\Omega)$. 
In the SU(2)-symmetric case, ${\cal H}_{0}$ and the Zeeman term 
$-\boldsymbol{B}\cdot\boldsymbol{S}_{\rm tot}$ commute with each other. 
As a result, the total spin exhibits just a 
simple precession motion around the $\boldsymbol{B}$ axis. 
This clearly indicates the absence of static growth of 
total magnetization. 
From this argument, we find that a magnetic anisotropy breaking 
the U(1) symmetry around the $\boldsymbol{B}$ axis is necessary 
to generate nonzero static magnetization. 
Notice that if the magnetic anisotropy conserves 
the U(1) symmetry about the $S^{z}$ axis, the mapping to the static system 
(\ref{eq:EffectiveHamiltonian}) is still valid. 
Fortunately, magnetic anisotropy generally exists 
in real magnetic materials. 

\begin{figure}[t]
\includegraphics[width=0.45\textwidth]{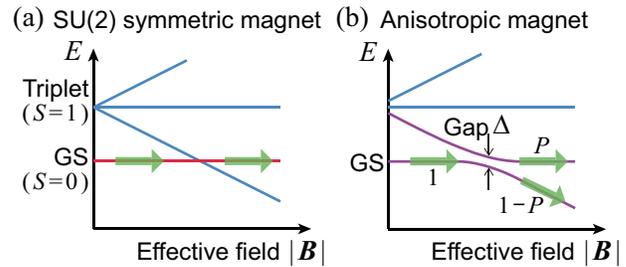}
\caption{(Color online) (a) Low-energy excitations as a function of 
the effective field $|\boldsymbol{B}|$ in a SU(2)-symmetric magnet 
${\cal H}_{0}$ and (b) in a magnet with magnetic anisotropy.}
\label{fig:EffModel}
\end{figure}

Based on a simple symmetry argument, let us qualitatively compare 
time evolution of quantum states in anisotropic and isotropic 
(SU(2)-symmetric) magnets, increasing the frequency $\Omega$. 
We assume that the GS of the SU(2)-symmetric magnet is paramagnetic 
with $S=0$. Usually, its lowest excitation is given by spin-$1$ 
triplet with $S=1$. 
If $\Omega$ (i.e., the effective field $|\boldsymbol{B}|$) is increased, 
the triplet excitations are split due to the Zeeman effect 
as shown in Fig.~\ref{fig:EffModel}(a). 
At a certain value of $\Omega$, a level crossing occurs 
between the GS and the lowest excitation of the triplet. 
However, if we start from the GS, the wave function is still 
equivalent to the GS even after the level crossing 
since there is no matrix element between the lowest two states 
due to the SU(2) symmetry. 
On the other hand, if we consider a magnet with magnetic anisotropy, 
the level crossing generally changes into an {\it avoided crossing} 
as shown in Fig.~\ref{fig:EffModel}(b). 
When the frequency $\Omega$ is monotonically increased, 
some weight of the wave function continuously follows the 
lowest-energy state and the total magnetization grows by unity, 
while the remaining weight tunnels into the upper state. 
This phenomenon would be described 
by the LZ tunneling picture~\cite{Landau32,Zener34}. 
With further increasing of $\Omega$, avoided crossings 
between the lowest two states successively take place and 
the magnetization would gradually grow. 
In order to make a larger weight of the state follow the GS of 
Eq.~(\ref{eq:EffectiveHamiltonian}), 
we have to suppress the LZ tunneling probability $P$ 
in Fig.~\ref{fig:EffModel}(b). From a theory of 
two-level LZ tunneling, the probability $P$ is known to 
rapidly decrease with decreasing the varying speed of $\Omega$. 
This argument based on the LZ tunneling clearly indicates that we should 
raise $\Omega$ as slowly as possible to realize magnetized states 
by the laser. Namely, an adiabatic change of $\Omega$ is ideal. 
Thanks to the recent development of laser technique, we can prepare 
lasers with gradually increasing its frequency $\Omega$. This technique 
is called {\it chirping}~\cite{SatoGonokami13,Kamada13}.

In conclusion, 
the argument in this section tells us that circular polarization, 
magnetic anisotropy, and chirping of laser would be significant for 
the realization of laser-induced magnetization process. 
Even if the magnetic anisotropy slightly breaks the U(1) symmetry, 
laser-induced magnetization is expected to occur.

\section{Laser-induced Magnetization Process}
\label{sec:LIMP}

The idea of the laser-induced magnetization process 
discussed in the previous section 
could be applied to general quantum magnets. 
In the remaining part of this paper, 
in order to quantitatively study the laser-induced dynamics, 
we concentrate on two simple but realistic one-dimensional (1D) 
spin-1/2 systems: Heisenberg antiferromagnetic (HAF) and 
ferro-ferro-antiferromagnetic (FFAF) models. 
The HAF model explains properties of many quasi-1D magnetic materials 
while the FFAF model exhibits a 1/3 magnetization plateau state 
with a SPT order as explained below, which 
describes $\rm Cu_{3}(P_{2}O_{6}OH)_{2}$~\cite{Hase06,Hase07,Hida94}.
The spin Hamiltonians are respectively given by 
\begin{align}
 {\cal H}_{\rm HAF}=&\sum_{j=1}^{L}J\boldsymbol{S}_{j}
   \cdot\boldsymbol{S}_{j+1},\nonumber\\
 {\cal H}_{\rm FFAF}=&\sum_{j=1}^{L/3}
      (-J_{\rm  F}\boldsymbol{S}_{3j-2}\cdot\boldsymbol{S}_{3j-1}
       -J_{\rm  F}\boldsymbol{S}_{3j-1}\cdot\boldsymbol{S}_{3j  }\nonumber\\
&\qquad+J_{\rm AF}\boldsymbol{S}_{3j  }\cdot\boldsymbol{S}_{3j+1}),\nonumber
\end{align}
where $\boldsymbol{S}_{j}$ is the spin-1/2 operator on site $j$ 
and $L$ is the system size. 
Coupling constants $J$, $J_{\rm F}$, and $J_{\rm AF}$ are all positive. 

If ${\cal H}_{0}$ is SU(2) symmetric, 
${\cal H}_{0}$ commute with $S_{\rm tot}^{x,y,z}$. 
In this case, magnetic (static) and laser (dynamic) parts 
in the time evolution operator are separated 
($e^{-i{\cal H}(t)t}=e^{-i{\cal H}_{0}t}
e^{i(A_{x}(t)S^{x}_{\rm tot}+A_{y}(t)S^{y}_{\rm tot})t}$), 
and the magnetization dynamics becomes trivial.
As we mentioned in Sec.~\ref{sec:BasicIdea}, we have to include 
a term that breaks SU(2) symmetry in ${\cal H}_{0}$ 
to lead to nonzero magnetization. 
As a small magnetic anisotropy term in ${\cal H}_{0}$, 
we here introduce a staggered Dzyaloshinsky-Moriya (DM) 
interaction~\cite{Dzyaloshinsky58}: 
\begin{equation}
 {\cal H}_{\rm DM}=\sum_{j=1}^{L}(-1)^{j}\boldsymbol{D}\cdot
   \boldsymbol{S}_{j}\times\boldsymbol{S}_{j+1}\nonumber
\end{equation}
with a DM vector $\boldsymbol{D}=(0,0,D)$ 
($D\ll J,J_{\rm AF},J_{\rm F}$). 
Such an anisotropy often appears in quasi-1D 
magnets~\cite{Dender97,Asano00,Feyerherm00,Kohgi01,Oshikawa97b,Sato04}. 
Note that any types of magnetic anisotropy 
such as Ising and single-ion anisotropies 
play the same role.

\begin{figure}[t]
\includegraphics[width=0.3\textwidth]{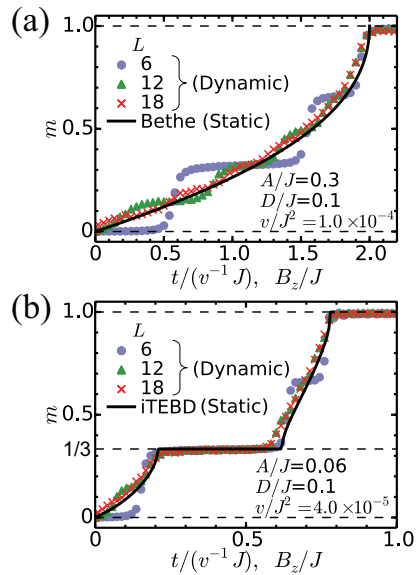}
\caption{(Color online) Laser-induced magnetization curves of 
(a) the HAF model and (b) the FFAF model with $J_{\rm F}=J_{\rm AF}=J$
realized by an upchirped circularly polarized laser in the slow chirp limit.
Solid lines are static magnetization curves 
from (\ref{eq:EffectiveHamiltonian})
obtained by Bethe ansatz and iTEBD ($D=B_{x}=0$). 
The horizontal axes for the dynamical and static process are
$t/(v^{-1}J)(=\Omega(t)/J)$ and $B_{z}/J$, respectively.}
\label{fig:MagCurve}
\end{figure}

We numerically study real time evolution 
of the wave functions in two models 
${\cal H}_{\rm HAF}$ and ${\cal H}_{\rm FFAF}$ 
with a staggered DM term under a circularly polarized laser. 
We start from the GS of ${\cal H}_{0}$, and then apply 
an upchirped circularly polarized laser. 
The pulse laser is modeled as follows. 
(i) Switch on: We first increase the amplitude from 0 to $A$ 
during $t=-10^{3}J^{-1}$ to 0.
During this process, $\Omega$ is zero. 
(ii) Chirping: We linearly increase the frequency as 
\begin{equation}
\Omega(t)=vt,\nonumber
\end{equation}
where $v$ is the chirping speed, 
i.e., the magnetic field of applied laser is 
represented as $(A_{x},A_{y})=A(\cos(vt^{2}/2),\sin(vt^{2}/2))$. 
In order to mimic an experimental situation,
we consider the case where the maximum field strength $A$
is small enough, e.g., $A/J, A/J_{\rm F,AF}\ll 1$,
while $\Omega$ can increase up to a large value 
such as $\Omega/J, \Omega/J_{\rm F,AF}\sim 1$. 
Though the increase of $A$ and $\Omega$ 
occurs simultaneously in real laser pulses, 
here we split them into two processes 
to understand the effect of chirping alone. 

We first calculate the GSs of ${\cal H}_{\rm HAF}+{\cal H}_{\rm DM}$ 
and ${\cal H}_{\rm FFAF}+{\cal H}_{\rm DM}$ 
by exact diagonalization in finite systems ($L=6,12,18$), 
and then numerically integrate the time-dependent Schr\"odinger equation 
$i\frac{d}{dt}|\Psi(t)\rangle={\cal H}(t)|\Psi(t)\rangle$ for 
the magnets under a circularly polarized laser, 
using the fifth order Runge-Kutta method. 
Hereafter, we use the normalized magnetization: 
\begin{equation}
 m\equiv M_{\rm tot}/M_{\rm tot}^{\rm st},\nonumber
\end{equation}
where $M_{\rm tot}\equiv \langle S_{\rm tot}^{z}\rangle$ 
and $M_{\rm tot}^{\rm st}$ is its saturated value. 

\begin{figure}[t]
\includegraphics[width=0.3\textwidth]{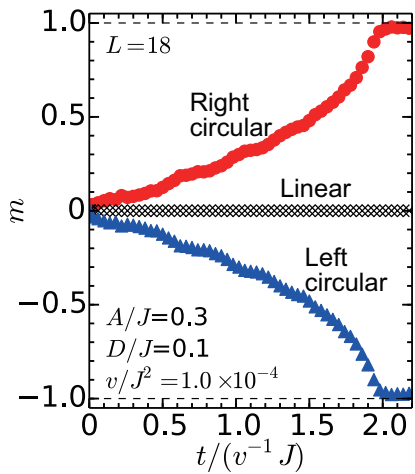}
\caption{(Color online) Laser-induced magnetization curves of the 1D 
HAF model with $L=18$ in right-circularly, left-circularly, 
and linearly polarized lasers. We have used the same parameters 
as those of Fig.~\ref{fig:MagCurve}(a).}
\label{fig:Polarization}
\end{figure}

Figure~\ref{fig:MagCurve}(a) shows the laser-induced magnetization curve 
of the HAF model plotted against the time. 
The chirping speed $v$ is set to a very small value. 
We will study the $v$ dependence later.
The curve is compared with the equilibrium one 
of the effective Hamiltonian (\ref{eq:EffectiveHamiltonian})
in static field $B_{z}$ with $B_{x}=D=0$ obtained 
by Bethe ansatz~\cite{Qin97,Cabra98}. 
Here we have ignored $B_{x}$ and $D$ in the static 
calculation since they are very small. 
We observe that the laser-induced magnetization curve 
converges to the static curve in the large $L$ limit. 
The FFAF model shows similar behaviors with 
one additional feature: the $m=1/3$ plateau state. 
As plotted in Fig.~\ref{fig:MagCurve}(b), with increasing $L$, 
the laser-induced magnetization curve also converges to 
the static curve obtained by 
infinite time evolving block decimation (iTEBD)~\cite{Vidal07} 
with matrix dimension $\chi=150$. 
During the process, the magnetization shows a plateau around 
$m=1/3$ with a width almost independent of the system size. 
We also confirmed the realization of 1/3 plateau 
for the same models with other anisotropies 
such as Ising-type and uniform DM interaction.

In addition to the FFAF model, 
we also numerically confirmed that a laser-induced dynamical 
1/3 plateau can be observed in a 1D $J_{1}$-$J_{2}$ spin 
model~\cite{Maeshima03,Okunishi03,Okunishi08,Hikihara10}. 
However, this system requires 
stronger DM interaction with $D\sim {\cal O}(J_{1,2})$ 
for generating the laser-driven plateau. 
This would be because the transition to the plateau phase is 
accompanied with a spontaneous breakdown of translational symmetry 
in the $J_{1}$-$J_{2}$ model (any symmetry breaking does not take place 
in the plateau of the FFAF model). A plateau with a broken symmetry 
is more fragile against perturbation than that with no symmetry breaking.

As we mentioned in Sec.~\ref{sec:BasicIdea}, 
if we change the sign of 
laser frequency from $\Omega$ to $-\Omega$, which is equivalent to 
changing the polarization of the laser, the magnetization is 
expected to point in the opposite direction. 
We confirm this prediction in the one-dimensional (1D) HAF model. 
Figure~\ref{fig:Polarization} shows numerical results for 
laser-induced magnetization curves of the HAF chain 
under right-circularly, left-circularly, and linearly 
polarized lasers with small chirping speed $v$. 
As expected, the magnetizations for right- and left-circularly 
polarized lasers are antiparallel with each other and 
linear polarization does not induce magnetization. 
In fact, we can prove rigorously that 
the size of induced magnetization is the same and 
the direction is opposite for right- and left-circularly 
polarized lasers applied to HAF and FFAF magnets 
with Ising, uniform DM, and staggered DM anisotropies 
(see Appendix~\ref{sec:Polarization}). 
It is also shown that the magnetization induced by 
linearly polarized laser is exactly zero. 
Although such a rigorous proof does not exist 
in general systems (e.g., systems in more than one dimension), 
it is expected that the magnetization points 
in the opposite direction if circular polarization is reversed. 
Moreover, a linearly polarized laser would not be useful 
for the magnetization growth.

\section{Landau-Zener tunneling}
\label{sec:LZtunneling}

\begin{figure}[t]
\includegraphics[width=0.3\textwidth]{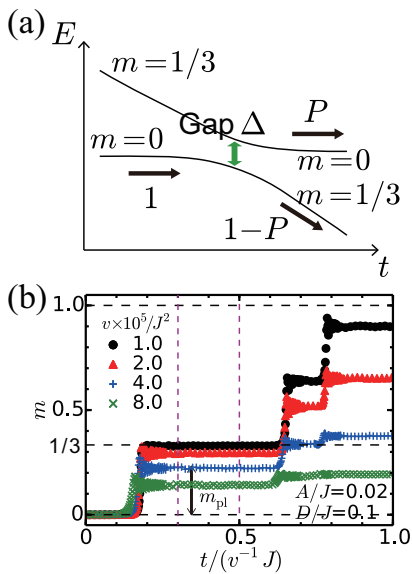}
\caption{(Color online) (a) Schematic picture of the LZ tunneling. 
(b) Laser-induced magnetization curves of the $L=6$ 
1D FFAF model for different
chirping speed $v$.
The height $m_{\rm pl}$ of the 1/3 plateau is defined as the average 
within $0.3\leq t/(v^{-1}J)\leq0.5$ (vertical dashed lines).}
\label{fig:ZenerTunnel}
\end{figure}

\begin{figure}[t]
\includegraphics[width=0.3\textwidth]{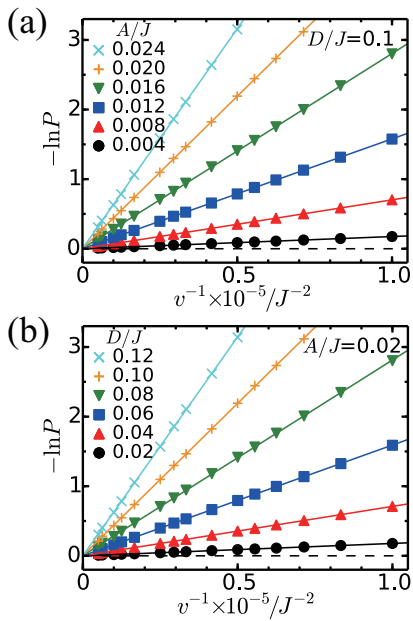}
\caption{(Color online) 
Plots of $-\ln P$ as a function of $v^{-1}$ 
(a) varying $A$ with fixed $D$ and (b) varying $D$ with fixed $A$. 
Solid lines show fittings of the data with 
a function $-\ln P\propto v^{-1}$.}
\label{fig:TunnelingProb}
\end{figure}

As shown in the previous section, if the chirping speed $v$ 
is sufficiently slow, 
it is possible to realize fully polarized magnetization 
as well as magnetization plateau states. 
Then, the natural question is 
``How slow should it be to see the plateau?'' 
It is crucial in finite width pulses. 

\begin{figure}[t]
\includegraphics[width=0.3\textwidth]{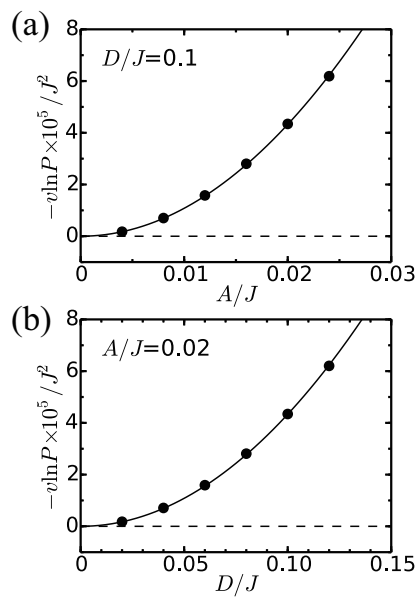}
\caption{(Color online) Scaling $-v\ln P$ ($\propto \Delta^{2}$) 
(a) as to $A$ with fixed $D$ and (b) as to $D$ with fixed $A$. 
Solid lines represent fittings with a function 
(a) $-v\ln P\propto A^{2}$ and (b)$-v\ln P\propto D^{2}$.}
\label{fig:GapScaling}
\end{figure}

To answer the question, we study the $v$ dependence of the 
magnetization curve. Figure~\ref{fig:ZenerTunnel}(b) shows that of 
the FFAF chain with $J_{\rm F}=J_{\rm AF}=J$ and $L=6$. 
We notice that the height of the $1/3$ plateau becomes 
lower as the speed becomes faster 
(note that other plateau-like states in Fig.~\ref{fig:ZenerTunnel}(b) 
are attributed to finite-size effect). 
This reduction can be explained by the LZ tunneling 
picture~\cite{Miyashita95}. 
When $v$ is sufficiently small, we can understand the magnetization 
dynamics from the effective static model~(\ref{eq:EffectiveHamiltonian}). 
As we increase $\Omega(=B_{z})$, 
the excited state with magnetization $m=1/3$ 
lowers its energy due to the Zeeman term 
$-\boldsymbol{B}\cdot\boldsymbol{S}_{\rm tot}$ 
and crosses the GS ($m=0$ sector) at a certain value of $\Omega$. 
(Strictly speaking, $S_{\rm tot}^{z}$, i.e., $m$, is not a good quantum number 
due to the presence of transverse field $A(=B_{x})$ term. 
However, we assume that $B_x$ is much smaller than $B_z=\Omega$ and 
$S_{\rm tot}^{z}$ can be approximated as a conserved quantity.)
Since $A$ and $D$ are finite and the spin-rotational symmetry is broken, 
the crossing becomes an avoided crossing, where we denote the gap between the 
$m=0$ and $m=1/3$ sectors as $\Delta$ (Fig.~\ref{fig:ZenerTunnel}(a)). 
In the present case of the $L=6$ FFAF model, 
this avoided crossing takes place at $t/(v^{-1}J)=\Omega/J\sim 0.174$. 
The wave function $|\Psi(t)\rangle$ is initially given by the GS ($m=0$).
After the LZ tunneling around the avoided crossing point, 
$|\Psi(t)\rangle$ becomes a linear superposition of 
the $m=1/3$ state (new GS) and the $m=0$ state (new excited state). 
Let us represent the probability of being in the new excited state and GS 
by $P$ and $1-P$, respectively. 
We can relate $P$ with the plateau height. 
To make this concrete, we define the averaged height $m_{\rm pl}$ 
as a mean value of $m$ within $0.3\leq \Omega/J\leq 0.5$ 
(Fig.~\ref{fig:ZenerTunnel}(b)). 
For the $1/3$ plateau, we can extract a relation
$m_{\rm pl}=(1-P)/3$. 
Then we can estimate $P$ from the numerically determined $m_{\rm pl}$. 
The LZ formula for the tunneling probability $P$ in a two-level 
system is known to be~\cite{Landau32,Zener34} 
\begin{equation}
 P=\exp(-c\Delta^{2}/v),
 \label{eq:TunnelProb}
\end{equation}
with a constant $c$. 
In order to verify whether this formula holds 
in our many-body problem, we numerically calculate $m_{\rm pl}$ for 
various values of $A$, $D$, and $v$. We plot $-\ln P$ as a function 
of $v^{-1}$ in Fig.~\ref{fig:TunnelingProb}. 
The data for fixed $A$ and $D$ are well fitted 
with a linear function (solid lines), which indicates 
the validity of Eq.~(\ref{eq:TunnelProb}). 
Figure~\ref{fig:GapScaling} shows that $-v\ln P$ is well 
fitted with $\propto A^{2}$ ($\propto D^{2}$) 
for fixed $D$ (A). Therefore, from the relation $-v\ln P=c\Delta^{2}$, 
the dependence of the tunneling gap on $A$ and $D$ is 
\begin{equation}
\Delta\propto AD.\label{eq:LZGap}
\end{equation}
At first glance, it may appear unexpected that 
the gap depends not only on the material parameter $D$
but also on the external field strength $A$. 
The most intuitive reasoning is that we need both
anisotropy and rotating magnetic field to induce a finite 
magnetization. Then Eq.~(\ref{eq:LZGap}) is just the leading order
in the series expansion. 
We can arrive at the same conclusion more systematically 
using a perturbation theory with respect to $D$ and $A$ 
(see Appendix~\ref{sec:LZgapPerturb} for details of the calculation). 

\begin{figure}[t]
\includegraphics[width=0.4\textwidth]{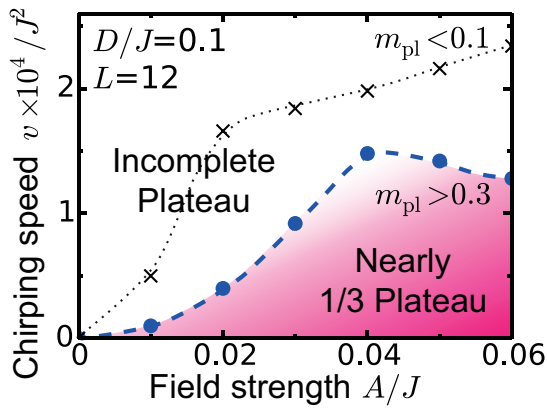}
\caption{(Color online) 
``Phase diagram'' of the laser-driven 1/3 plateau state of 
the FFAF model with $J_{\rm F}=J_{\rm AF}=J$ 
in the parameter space $(A,v)$. 
The dashed and dotted lines 
are contours of plateau height $m_{\rm pl}=0.3$ and 0.1, respectively. 
This tells us the condition for realizing an ideal plateau state 
with large magnetization.} 
\label{fig:PhaseDiag}
\end{figure}

In Fig.~\ref{fig:PhaseDiag}, 
we show a ``phase diagram'' for realizing the plateau state 
in the parameter space of field strength $A$ and chirping speed $v$ 
for the 1D FFAF model with $J_{\rm F}=J_{\rm AF}=J$ and $L=12$. 
We represent contours of plateau height $m_{\rm pl}=0.3$ and 0.1 
by dashed and dotted lines, respectively. 
The phase diagram clearly indicates that 
strong laser with slow chirping speed 
(the bottom right part in Fig.~\ref{fig:PhaseDiag}) is preferred 
for achieving a magnetization plateau with large magnetization. 
We define a ``nearly 1/3 plateau'' region as $m_{\rm pl}>0.3$
(lower than the dashed line, colored with shade). 
This region has a domelike structure with a peak near $A/J\sim 0.04$.
On the left side of the peak ($A/J<0.04$), the boundary is parabolic in $A$, 
which can be understood by the LZ tunneling.
Using Eqs.~(\ref{eq:TunnelProb}) and (\ref{eq:LZGap}), 
the criterion for a sharp laser-induced plateau 
is ($\alpha$: nonuniversal constant)
\begin{equation}
 v/A^{2}<\alpha D^{2},\label{eq:criteria}
\end{equation}
which explains the parabolic feature of the contour $m_{\rm pl}=0.3$. 
To make connection with experiment, 
we note that if we assume $J=10$ meV, $A=0.4$ meV $\sim 4$ T, and $D=1$ meV, 
then $v\sim10^{-4}J^{2}\sim 6\times 10^{-4}$ THz/ps 
is necessary for Eq.~(\ref{eq:criteria}). 
We also mention that other magnetic anisotropies will play the same role
as the DM interaction $D$ in Eq.~(\ref{eq:criteria}). 

\begin{figure}[t]
\includegraphics[width=0.3\textwidth]{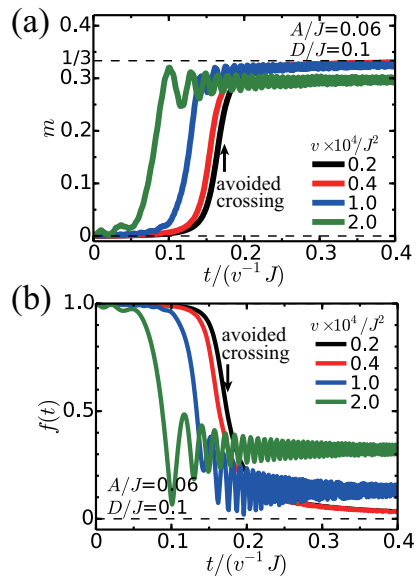}
\caption{(Color online) Time evolution of (a) magnetization $m$ and 
(b) overlap $f(t)$ between a wave function 
at time $t$ and the initial state 
for various chirping speed $v$. 
Arrows represent the time when 
the system goes through an avoided crossing 
($t/(v^{-1}J)\sim 0.174$).} 
\label{fig:Tunnel}
\end{figure}

In the strong field side ($A/J>0.04$), 
the contour $m_{\rm pl}=0.3$ starts to bend down. 
It implies that this region cannot be understood by the LZ formula. 
To see how the LZ theory breaks down, 
we perform numerical simulations at $A/J=0.06$ 
for various chirping speeds $v$. 
In addition to the magnetization $m$, we calculate the 
overlap $f(t)$ between the wave function $|\Psi(t)\rangle$ 
at time $t$ and the initial state $|\Psi(0)\rangle$: 
\begin{equation}
 f(t)\equiv|\langle\Psi(0)|\Psi(t)\rangle|^{2}.\nonumber
\end{equation}
The results are presented in Fig.~\ref{fig:Tunnel}. 
For small $v$, the increase of $m$ 
and the drop of $f(t)$ happens 
around $t/(v^{-1}J)\sim 0.174$, 
indicated by arrows in Fig.~\ref{fig:Tunnel}. 
The corresponding frequency $\Omega/J\sim 0.174$ 
is almost equal to the point of the minimum gap 
in Fig.~\ref{fig:ZenerTunnel}(a). Therefore, for small $v$, 
the LZ tunneling picture is valid and the transition 
from the initial state ($m=0$) to the plateau state ($m=1/3$) 
instantly happens around the single point $t/(v^{-1}J)\sim 0.174$. 
However, when the chirping speed becomes fast, 
i.e., larger $v$, this picture starts to fail. 
The increase of $m$ and drop of $f(t)$ start earlier than 
$t/(v^{-1}J)\sim 0.174$, and both quantities show oscillations. 
The LZ theory assumes that the tunneling happens 
at a single point where the energy gap is smallest, 
that is, the avoided crossing point 
(see Fig.~\ref{fig:ZenerTunnel}(a)). 
In the region of $A/J>0.04$, 
this necessary condition for the LZ theory is violated. 
It is presumably because a direct hybridization 
between the initial ($m=0$) and plateau ($m=1/3$) states 
begins before the LZ tunneling takes place 
due to the strong laser amplitude $A/J$. 
An oscillation seen in $m$ and $f(t)$, supposedly 
caused by a difference of phase factor 
between the initial and plateau states, 
supports our speculation. 
In addition, a hybridization is concerned with not only 
these two states but also other states with higher energy, 
which makes the LZ picture even worse. 
In the result, 
a crossover from the LZ mechanism in the weak field side 
($A/J<0.04$) to a direct hybridization in the strong field 
side ($A/J>0.04$) leads to the domelike structure.

\section{Plateau with SPT order}
\label{sec:Topological}

\begin{figure}[t]
\includegraphics[width=0.45\textwidth]{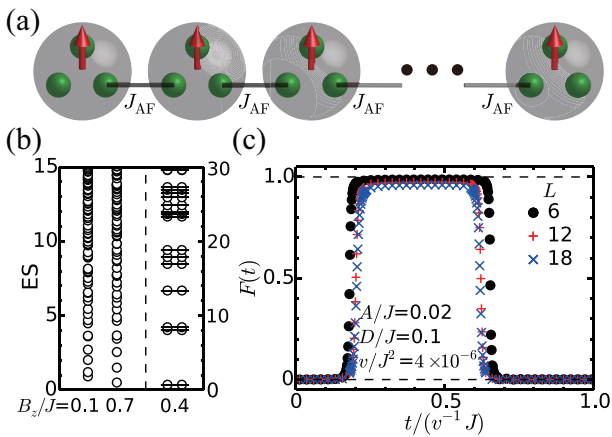}
\caption{(Color online) 
(a) Schematic VBS picture of the static 1/3 plateau phase in 
the FFAF chain. (b) ES of the static magnetized states 
in the FFAF chain with $D=0$ calculated by iTEBD ($\chi=150$). 
The system is partitioned at an antiferromagnetic bond. 
(c) Fidelity of the laser-induced dynamical state 
with respect to the static plateau with the SPT order.}
\label{fig:Topological}
\end{figure}

Finally, we discuss topological properties of the dynamically 
induced $1/3$ plateau state in the 1D FFAF model. 
First, we study the GS of the FFAF model 
in static field $B_{z}$ using iTEBD. 
The SPT order in spin systems can be detected by 
the entanglement spectrum (ES)~\cite{Pollmann10,Gu09}. 
To define ES for a quantum state $|\psi\rangle$, 
we divide the system into two subspaces A and B. 
From the Schmidt decomposition 
$|\psi\rangle=\sum_{i}\lambda_{i}
|\psi_{\rm A}\rangle_{i}\otimes|\psi_{\rm B}\rangle_{i}$,
the ES is defined as $-\ln(\lambda_{i}^{2})$ $(i=1,\ldots,\chi)$, 
where $\lambda_{i}$ is normalized as $\sum_{i}\lambda_{i}^{2}=1$. 
In the FFAF model, we evaluate the ES by 
cutting the system at the antiferromagnetic bond $J_{\rm AF}$, 
and the result is plotted in Fig.~\ref{fig:Topological}(b). 
We see that the ES is doubly degenerate in the plateau state 
at $B_{z}/J=0.4$, but there is no degeneracy 
at $B_{z}/J=0.1$ and 0.7. 
This degeneracy is a clear signature of the SPT order 
in the $1/3$ plateau state. In fact, we can relate this state 
with a valence bond solid (VBS) state~\cite{Affleck87} as follows. 
As schematically shown in Fig.~\ref{fig:Topological}(a),
in the limit of $J_{\rm F}\gg J_{\rm AF}$, 
ferromagnetically coupled three adjacent $S=1/2$ spins can be 
regarded a ``site'' with $S=3/2$, and neighboring trimers 
are coupled by an antiferomagnetic bond $J_{\rm AF}$. 
In the 1/3 plateau state, one of the three spin-1/2 components
is fully polarized (represented by an arrow), 
and the remaining two spins form singlet pairs (solid line) 
with their neighbors. If we assume that this VBS picture of 
the plateau state survives up to $J_{\rm F}\sim J_{\rm AF}$, 
the plateau must possess the same topological nature as the Haldane state 
in the $S=1$ HAF model~\cite{Affleck87,Haldane83}. This argument clearly 
explains the double degeneracy of the ES in the plateau state.
The symmetry protection of magnetization plateau states 
will be discussed in detail elsewhere~\cite{TakayoshiTBA}.

We next turn to the laser-induced plateau state. 
How close is the dynamical plateau state 
to the static SPT state? 
To answer this question, we calculate the 
fidelity (overlap): 
\begin{equation}
 F(t)\equiv\big|\langle\Phi_{\rm topo}^{\rm st}|
   \Psi(t)\rangle\big|^{2}\nonumber
\end{equation}
by numerical diagonalization, 
where $|\Phi_{\rm topo}^{\rm st}\rangle$ represents 
the static GS of the FFAF chain in $B_{z}/J=0.4$ ($D=A=0$) and 
$|\Psi(t)\rangle$ is the laser-induced state. 
Figure~\ref{fig:Topological}(c) shows that 
the overlap $F(t)$ is larger than 95 \% when the dynamical state 
$|\Psi(t)\rangle$ is in the plateau phase ($0.2\lesssim B_{z}/J\lesssim 0.6$) 
while $F(t)\sim 0$ when $|\Psi(t)\rangle$ 
is outside the plateau phase. 
This result is almost independent of the system size $L$. 
Thus, we conclude that if the chirping speed $v$ is sufficiently small,
the SPT phase can be achieved dynamically. 
In a rigorous sense, we have to switch off the laser amplitude 
or neglect the DM anisotropy 
in order to protect the topological properties of the dynamical 
plateau state~\cite{Pollmann10}. 
Such a subtlety can be avoided in the case of Ising anisotropy.

\section{Conclusion and Discussions}
\label{sec:Discuss}

We have proposed a novel scheme of 
``laser-induced magnetization curves'' 
for general quantum magnets. 
An upchirped circularly polarized laser in the 
THz regime is required, and the material must have
small but finite magnetic anisotropy. 
From our numerical calculation in realistic spin models,
the criterion for laser strength and chirping speed was
obtained, which is explained by the LZ picture.  
Using this method, it is even possible to 
realize a SPT phase dynamically. 
In this paper, we have demonstrated 
laser-induced magnetization curves 
in two types of magnets: HAF and FFAF chains. 
However, as can be seen from our analyses, 
this phenomenon does not exploit specific properties of these magnets. 
Therefore, we again stress that laser-induced magnetization 
curves would be attained by laser chirping in a wide variety of 
magnetic materials in any dimension. 
We also note that the direction of the magnetization growth can be changed 
by switching the helicity of the laser. 

In addition to the above necessary condition for the magnetization dynamics, 
let us briefly discuss some desired conditions for easily realizing a higher 
value of laser-induced magnetization. 
Large magnetic anisotropies are desired to open a wide gap $\Delta$ 
in Fig.~\ref{fig:ZenerTunnel}(a). Magnets with 
large spin magnitude $S$ are preferred since the value $S$ is regarded 
as the coupling constant of the Zeeman term, 
namely a large value $S$ helps the laser amplitude to effectively increase. 
When the Zeeman coupling is too small, effects of the laser might be 
masked by several noises such as thermal fluctuation, dipole interaction, 
spin-phonon coupling, and so on. 
Magnets with ferromagnetic exchanges are also favorable since a 
ferromagnetic interaction generally decreases the value of saturation field 
and it reduces the necessary time of chirping. 
The time interval of chirping needed for full magnetization curves 
in Fig.~\ref{fig:MagCurve} is estimated to be about 
$2.0\times 10^{4}J^{-1}$ for both HAF and FFAF chains. 
Assuming $J\sim 10$ meV ($J^{-1}\sim 0.4$ ps), 
it amounts to $2.0\times 10^{4}J^{-1}\sim 8$ ns. 
It is generally difficult to generate an ideal upchirped laser 
for a long time interval in real experiments. 
However, even a nonideal laser is expected to 
generate a finite magnetization if the laser frequency is comparable to 
the exchange coupling of the target materials~\cite{Takayoshi13}. 
In an actual experimental setup, 
heating caused by laser would smear the plateau structure. 
In order to prevent such heating, 
we should choose the laser frequency far from 
any resonant frequencies of phonons, magnons, etc.: 
excitations which thermalize the sample. 
Furthermore, if the LZ gap is large enough 
(thanks to, e.g., large Zeeman coupling), 
a laser would induce a finite magnetization before 
the temperature considerably increases.

A pump-probe experiment with two laser pulses would be one of the 
simple realistic ways of measuring laser-induced magnetizations. 
We prepare a circularly polarized laser pulse to coherently control 
the magnetization of a target material. The other pulse is used 
to measure the magnetization, for example, 
through the inverse Faraday or Kerr effect. 
Notice that the laser-induced magnetization curve differs 
considerably from standard magnetic resonance phenomena. 
Our scheme does not have a specific energy scale 
concerning the laser frequency $\Omega$ 
(although $\Omega$ should be the order of exchange couplings) 
while a magnetic resonance usually happens (or becomes strong) 
at a certain frequency characteristic of the magnet.

Effects of dissipation from an environment and 
heating magnets by laser, which are not considered here, 
are important future problems. 
In this paper, to see the laser-driven dynamics of magnets, 
we have focused on finite-size systems. Investigation on the effects 
of laser in the thermodynamic limit is another important issue. 

We finally comment on a possible combination of the 
current method and actual static magnetic fields. 
Namely, a static Zeeman term $B_{\rm st}S_{\rm tot}^{z}$ can be added 
to ${\cal H}_{0}$, and after the mapping from 
Eq.~(\ref{eq:HamiltonianWithLaser}) to Eq.~(\ref{eq:EffectiveHamiltonian}),
the total effective field becomes 
\begin{equation}
B^{\rm total}_{z}=\Omega+B_{\rm st}.
\end{equation}
This fact would be practically useful in experiments. 
For example, we can study the destruction and recovery of 
the topological state starting from the plateau state 
in finite static fields and then applying 
a pulse circularly polarized laser. 

\acknowledgements
We acknowledge Hideki Hirori, Kazuhiko Misawa, 
and Koichiro Tanaka for illuminating discussions 
on THz laser experiments. 
This work is supported by Grants-in-Aid from JSPS, 
Grant No. 25287088, No. 26870559 (M.S.), and No.23740260 (T.O.).

\appendix

\section{Relation between laser polarization and induced magnetization}
\label{sec:Polarization}

Magnetization $m(>0)$ along the $z$ axis is induced 
when a right-circularly polarized laser 
$A(\cos(\Omega t),\sin(\Omega t))$ is applied to 
HAF and FFAF magnets with anisotropy. 
In the following, we prove that 
the induced magnetization becomes $-m$ 
for an application of a left-circularly polarized laser 
$A(\cos(\Omega t),-\sin(\Omega t))$ 
in the case of HAF and FFAF magnets 
with anisotropies of 
(i) Ising, (ii) uniform DM, and (iii) staggered DM types. 
In addition, we can show that magnetization 
along the $z$ axis is exactly zero ($m=0$) 
for a linearly polarized laser $A(\cos(\Omega t),0)$. 
We write the time-dependent Schr\"odinger equation 
for the right-circularly (left-circularly) polarized laser as 
\begin{equation}
 i\partial_{t}|\Psi_{\rm R(L)}(t)\rangle=
   {\cal H}_{\rm R(L)}|\Psi_{\rm R(L)}(t)\rangle.\nonumber
\end{equation}
The initial states are the same, 
$|\Psi_{\rm R}(0)\rangle=|\Psi_{\rm L}(0)\rangle$. 

(i) Ising anisotropy: 
Let us consider $\pi$ rotation around the $x$ axis, $\pi_{x}$. 
This rotation $\pi_{x}$ changes $S_{\rm tot}^z$ to $-S_{\rm tot}^z$ 
and reverses the circular polarization from right to left in the Zeeman term. 
On the other hand, it keeps ${\cal H}_{\rm HAF}$, ${\cal H}_{\rm FFAF}$, and 
Ising anisotropy $J_{z}\sum_{j=1}^{L}S_{j}^{z}S_{j+1}^{z}$ unchanged. 
Thus, ${\cal H}_{\rm L}=\pi_{x}{\cal H}_{\rm R}\pi_{x}^{-1}$. 
Since the initial state is invariant under $\pi_{x}$ 
($\pi_{x}|\Psi_{\rm R}(0)\rangle=|\Psi_{\rm R}(0)\rangle$), 
$|\Psi_{\rm L}(t)\rangle$ is obtained by 
$\pi_{x}|\Psi_{\rm R}(t)\rangle$. 
Therefore, if magnetization $m$ is induced 
for a right circularly polarized laser, 
$-m$ is induced for the case of left circular polarization. 
For a linear polarization, 
the system is invariant under $\pi_{x}$. 
Then, $m$ is exactly zero due to $m=-m$. 

(ii) Uniform DM anisotropy: 
We represent the inversion of the chain as $I$, 
i.e., the site $j$ corresponds to $L+1-j$. 
The succession of $\pi_{x}$ and $I$ does not 
change ${\cal H}_{\rm HAF}$, ${\cal H}_{\rm FFAF}$, 
and uniform DM anisotropy 
$D_{\rm u}\sum_{j=1}^{L}(S_{j}^{x}S_{j+1}^{y}-S_{j}^{y}S_{j+1}^{x})$ 
while the circular polarization is reversed. 
From the same logic as in (i), 
$|\Psi_{\rm L}(t)\rangle=I\pi_{x}|\Psi_{\rm R}(t)\rangle$. 
Therefore, $m\to -m$ if the circular polarization is changed 
from right to left and $m=0$ for a linear polarization. 

(iii) Staggered DM anisotropy: 
We consider one (three) site translation of the system $T$ ($T^{3}$), 
i.e., the site $j$ corresponds to $j+1$ ($j+3$). 
The succession of $\pi_{x}$ and $T$ ($T^{3}$) does not 
change ${\cal H}_{\rm HAF}$ (${\cal H}_{\rm FFAF}$) and 
staggered DM anisotropy 
$D_{\rm s}\sum_{j=1}^{L}(-1)^{j}(S_{j}^{x}S_{j+1}^{y}-S_{j}^{y}S_{j+1}^{x})$ 
while the circular polarization is reversed. 
Thus, the proof is the same as in (ii). 

\section{Landau-Zener gap in dynamical magnetization curves}
\label{sec:LZgapPerturb}

We explain the reason why the LZ gap 
for a magnetization step is proportional to $AD$ 
in the FFAF model (see Eq.~(\ref{eq:LZGap})) 
using perturbation theory with respect to $A$ and $D$. 
Following the main text, we focus on the finite system with $L=6$. 

Through the unitary transform $U=\exp(i\Omega S_{\rm tot}^{z} t)$, 
the effective static Hamiltonian is mapped to
\begin{align}
 {\cal H}_{\rm eff}=&{\cal H}_{\rm FFAF}
   +D\sum_{j}(-1)^{j}(S_{j}^{x}S_{j+1}^{y}
                     -S_{j}^{y}S_{j+1}^{x})\nonumber\\
   &\qquad\quad-\Omega S^{z}_{\rm tot}-AS^{x}_{\rm tot}.\quad 
 \label{eq:EffectiveModel}
\end{align}
In the limit of $A\to 0$, the $z$ component of total spin $S^{z}_{\rm tot}$ 
is a good quantum number, and we can define the lowest energy states in 
${S}_{\rm tot}^{z}=0$ ($m=0$) and $S_{\rm tot}^{z}=1$ ($m=1/3$) sectors as 
$|\psi_{0}\rangle$ and $|\psi_{1/3}\rangle$, respectively. 
These two states come close to each other by tuning $\Omega$. 

Let us further perform a spin rotation around the $y$ axis 
so the magnetic field $(A,0,\Omega)$ becomes parallel to the $z$ axis. 
The corresponding unitary operator is $U_{y}=e^{i\theta S_{\rm tot}^{y}}$
with $\cos\theta=\Omega/\sqrt{\Omega^{2}+A^{2}}$ and 
$\sin\theta=A/\sqrt{\Omega^{2}+A^{2}}$. We note that 
the angle $\theta$ is very small ($|\theta|\ll 1$) since 
the laser amplitude $A$ is usually much smaller than the frequency $\Omega$, 
i.e., $|A|\ll \Omega$. Via this spin rotation, new spin operators 
are given as  
\begin{align}
\begin{pmatrix}
\tilde{S}^{x}\\ \tilde{S}^{z}
\end{pmatrix}
\equiv U_{y}
\begin{pmatrix}
S^{x}\\ S^{z}
\end{pmatrix}
U_{y}^{-1}=
\begin{pmatrix}
\cos\theta&\sin\theta\\
-\sin\theta&\cos\theta
\end{pmatrix}
\begin{pmatrix}
S^{x}\\ S^{z}
\end{pmatrix}.\nonumber
\end{align}
Similarly, the Hamiltonian is transformed into 
\begin{align}
 \tilde{\cal H}_{\rm eff}=
   U_{y}{\cal H}_{\rm eff}U_{y}^{-1}={\cal H}_{\rm FFAF}+{\cal H}_{\rm D}
-\sqrt{\Omega^{2}+A^{2}}\tilde{S}^{z}_{\rm tot}\nonumber
\end{align}
while $\tilde{S}^{y}=S^{y}$.
The second term $(\propto D)$ is given by
\begin{align}
 &{\cal H}_{\rm D}=D({\cal H}_{\rm cos}+{\cal H}_{\rm sin}),\nonumber\\
 &{\cal H}_{\rm cos}=\cos\theta
   \sum_{j}(-1)^{j}(\tilde{S}_{j}^{x}\tilde{S}_{j+1}^{y}
                   -\tilde{S}_{j}^{y}\tilde{S}_{j+1}^{x}),\nonumber\\
 &{\cal H}_{\rm sin}=-\sin\theta
   \sum_{j}(-1)^{j}(\tilde{S}_{j}^{y}\tilde{S}_{j+1}^{z}
                   -\tilde{S}_{j}^{z}\tilde{S}_{j+1}^{y}).\nonumber
\end{align}
Note that the form of ${\cal H}_{\rm FFAF}$ is invariant under 
the rotation of $U_{y}=e^{i\theta S^{y}_{\rm tot}}$ due to 
the SU(2) symmetry. 
Here we introduce the normalized magnetization per site 
of the model $\tilde{\cal H}_{\rm eff}$ 
as $\tilde{m}=2\langle \tilde{S}^{z}_{\rm tot}\rangle/L$.
In the case of $D=0$, $\tilde S^{z}_{\rm tot}$ is a good quantum number, 
and therefore we can define the ground states of 
$\tilde{S}^{z}_{\rm tot}=0$ ($\tilde m=0$) and $1$ ($1/3$) sectors as 
$|\tilde\psi_0\rangle$ and $|\tilde\psi_{1/3}\rangle$, respectively. 
Since $|A|\ll \Omega$ and the angle $\theta$ is very small, 
the two states $|\tilde\psi_{0}\rangle$ and $|\tilde\psi_{1/3}\rangle$
are respectively very close to $|\psi_{0}\rangle$ and $|\psi_{1/3}\rangle$. 

For the effective Hamiltonian $\tilde{\cal H}_{\rm eff}$, 
let us treat the second term ${\cal H}_{\rm D}$ as the perturbation. 
When $D=0$, the energy levels of $|\tilde{\psi}_{0}\rangle$ and 
$|\tilde{\psi}_{1/3}\rangle$ cross at a certain point, 
$\Omega=\Omega_{\rm c}$. However, this degeneracy is lifted and 
the level crossing becomes a level repulsion 
when $D$ is introduced. This gap of the level repulsion is nothing but 
the LZ gap. Let us study the effect of finite $D$ around 
the degenerate point $\Omega=\Omega_{\rm c}$. 
Applying the perturbation theory to the degenerate states 
$|\tilde{\psi}_{0}\rangle$ and $|\tilde{\psi}_{1/3}\rangle$, 
the first-order perturbation energy $\epsilon_{1}$ 
from ${\cal H}_{\rm D}$ is given as 
a solution of the following eigenvalue problem: 
\begin{equation}
\det
\begin{pmatrix}
\displaystyle
 \langle\tilde\psi_{0}|{\cal H}_{\rm D}
     |\tilde\psi_{0}\rangle-\epsilon_{1}&
\displaystyle
 \langle\tilde\psi_{0}|{\cal H}_{\rm D}
     |\tilde\psi_{1/3}\rangle\\
\displaystyle
 \langle\tilde\psi_{1/3}|{\cal H}_{\rm D}
     |\tilde\psi_{0}\rangle&
\displaystyle
\langle\tilde\psi_{1/3}|{\cal H}_{\rm D}
|\tilde\psi_{1/3}\rangle-\epsilon_{1}
\end{pmatrix}
=0.\label{eq:EigenPerturb}
\end{equation}

In the case of $D\neq 0$ and $A=0$, 
$|\tilde\psi_{0,1/3}\rangle$ is reduced to $|\psi_{0,1/3}\rangle$ 
and ${\cal H}_{\rm D}$ is equal to the original staggered DM interaction, 
namely, ${\cal H}_{\rm D}\to D{\cal H}_{\rm cos}$ with $\cos\theta=1$. 
The DM interaction ${\cal H}_{\rm cos}$ commutes with $\tilde S_{\rm tot}^z$
and it does not change the value of $\tilde S_{\rm tot}^z$. 
This leads to 
\begin{equation}
 \langle\tilde\psi_{0}|(\tilde{S}_{j}^{x}\tilde{S}_{j+1}^{y}
                 -\tilde{S}_{j}^{y}\tilde{S}_{j+1}^{x})
|\tilde\psi_{1/3}\rangle=0.
 \nonumber
\end{equation}
Therefore, the off-diagonal term of Eq.~(\ref{eq:EigenPerturb}) is zero 
and only the diagonal matrix elements 
$\langle\tilde\psi_{0,1/3}|{\cal H}_{\rm D}|\tilde\psi_{0,1/3}\rangle$ 
can be finite when $A=0$. 
However, if we suitably tune the external field (frequency) 
from $\Omega_{\rm c}$ to $\Omega_{\rm c}+\delta\Omega$ ($\delta\Omega\propto D$), 
we can remove the diagonal elements. 
Namely, the level crossing survives even after introducing 
the $D$ term and off-diagonal elements are necessary to 
generate a LZ gap. 

Next, we add a finite transverse field $A$ at the new level crossing point 
$\Omega=\Omega_{\rm c}+\delta\Omega$. 
In this case, ${\cal H}_{\rm sin}$ appears 
in the perturbation part ${\cal H}_{\rm D}$.
Since 
$\tilde{S}_{j}^{y}\tilde{S}_{j+1}^{z}-\tilde{S}_{j}^{z}\tilde{S}_{j+1}^{y}$ 
changes $\tilde{S}_{\rm tot}^{z}$ by $\pm 1$, we have 
\begin{align}
 &\langle\tilde\psi_{0}|
   (\tilde{S}_{j}^{y}\tilde{S}_{j+1}^{z}
   -\tilde{S}_{j}^{z}\tilde{S}_{j+1}^{y})|\tilde\psi_{0}\rangle=0\nonumber\\
 &\langle\tilde\psi_{1/3}|(
   \tilde{S}_{j}^{y}\tilde{S}_{j+1}^{z}
  -\tilde{S}_{j}^{z}\tilde{S}_{j+1}^{y})|\tilde\psi_{1/3}\rangle=0.\nonumber
\end{align} 
Thus, the eigenvalue equation in the subspace of $|\tilde\psi_{0,1/3}\rangle$ 
is expressed as 
\begin{equation}
\det
\begin{pmatrix}
\displaystyle
-\epsilon_{1}&
\displaystyle
 D\langle\tilde\psi_{0}|{\cal H}_{\rm sin}
     |\tilde\psi_{1/3}\rangle\\
\displaystyle
 D\langle\tilde\psi_{1/3}|{\cal H}_{\rm sin}
     |\tilde\psi_{0}\rangle&
\displaystyle
-\epsilon_{1}
\end{pmatrix}
=0.\label{eq:EigenPerturb2}
\end{equation}
Since we can make approximations $\cos\theta\approx 1$ 
and $\sin\theta\approx \tan\theta= A/\Omega$ 
due to $A\ll \Omega$, 
the off-diagonal terms in Eq.~(\ref{eq:EigenPerturb2}) are 
proportional to $AD/\Omega$. Therefore, the $D$-induced LZ gap 
$\tilde\Delta$ is proportional to $AD$ up to the leading order of $A$ and $D$. 
Even if we consider the LZ gap from the standpoint of ${\cal H}_{\rm eff}$ 
instead of $\tilde{\cal H}_{\rm eff}$, 
the $D$-induced gap $\Delta$ in the space of 
$|\psi_{0,1/3}\rangle$ is also proportional to $AD$ 
because $|\tilde{\psi}_{0,1/3}\rangle$ can be approximately equal to 
$|\psi_{0,1/3}\rangle$. 

The above argument based on the perturbation theory can also be applied to 
systems with other magnetic anisotropies with the U(1) symmetry 
around the $z$ axis (e.g., Ising anisotropy 
${\cal H}_{\rm Ising}=J_{z}\sum_{j}S_{j}^{z}S_{j+1}^{z}$ 
or uniform DM interaction).

\end{document}